\documentclass[RNAAS]{aastex62}

\usepackage{lmodern}
\usepackage{graphicx} 
\usepackage{textcomp}
\usepackage{amsmath}
\usepackage[english]{babel}
\usepackage{float}
\usepackage{hyperref}
\usepackage{amssymb}
\usepackage{xcolor}
\usepackage{aas_macros}
\usepackage{array}   

\newcolumntype{L}{>{$}l<{$}}
\newcolumntype{C}{>{$}c<{$}}

\newcommand{\dl}{\texttt{diskline}}
\newcommand{\sd}{\texttt{shaddisk}}
\newcommand{\kec}{\texttt{kerrconv}}
\newcommand{\ked}{\texttt{kerrdisk}}
\newcommand{\rb}{\texttt{rdblur}}
\newcommand{\rl}{\texttt{relline}}
\newcommand{\rc}{\texttt{relconv}}
\newcommand{\gauss}{\texttt{*gaussian}}
\newcommand{\xspec}{\textsc{Xspec}}

\newcommand{\fekal}{Fe K$\alpha$}

\newcommand{\conln}{\textit{ConvolutionInLnSpace}}

\begin{document}
	
\title{A note on some discrepancies in convolution models in X-ray spectral analysis}
\author{R. La Placa}	
\affiliation{INAF--Osservatorio Astronomico di Roma, via Frascati 33, 00040 Monteporzio Catone (Roma), Italy}
\affiliation{Dipartimento di Fisica, Universit\`a di Roma ``Tor Vergata'', Via della Ricerca Scientifica, 00133 Roma, Italy}
\affiliation{Dipartimento di Fisica, Universit\`a La Sapienza, Piazzale Aldo Moro 5, 00185 Roma, Italy}

\author{A. F. Gambino}
\affiliation{Universit\`a degli Studi di Palermo, Dipartimento di Fisica e Chimica - Emilio Segr\`e, via Archirafi 36 - 90123 Palermo, Italy}
\affiliation{INAF-IASF Palermo, Via U. La Malfa 153, I-90146 Palermo, Italy}

%
%
%
%

\begin{abstract}

Convolution models are powerful tools in many fields of spectral and image analysis owing to their wide applicability, and X-ray astrophysical spectral analysis is no exception. We found that relativistically broadened \fekal\ line profiles obtained through many convolution models both within and without \xspec\ show deviations from the profiles produced by their non-convolution counterparts. These discrepancies depend on the energy grid considered and on the shape of both the kernel and the underlying spectrum, but can reach as high as 10\% of the flux in certain energy bins. We believe that this effect should be taken into consideration, considering how often these models are used to study spectral features of lower relative intensity, and advise great discretion in using them. 

\end{abstract}

\section{} \label{sec:BatOutOfHell}


Convolution models allow to apply modifications due to various physical processes to any spectral feature under study: in general, a kernel function describing the shape resulting from a certain physical effect at a fixed energy or condition is convolved over an underlying spectrum to apply the same physical effect over the whole energy range considered. The intrinsically wide applicability of these models makes them extremely practical and they are thus used ubiquitously in spectral analysis. Recently, however, we found differences in \fekal\ line profiles generated with standard line profile models and their convolution counterparts.

In particular, we simulated a high-quality spectrum within \xspec, the most widely used data analysis program for X-ray spectra, using \sd\ as a local model to provide the \fekal\ line profile. We then substituted it before fitting with the convolution version of the model, \texttt{shadconv}, applied to a 
Gaussian profile with null sigma (essentially a Dirac delta). To our surprise, some residuals were found that showed high statistical significance even though the model parameters were exactly the same. 

\begin{figure}[h]
	\centering
	\gridline{
		\hspace{-1.cm}
		\fig{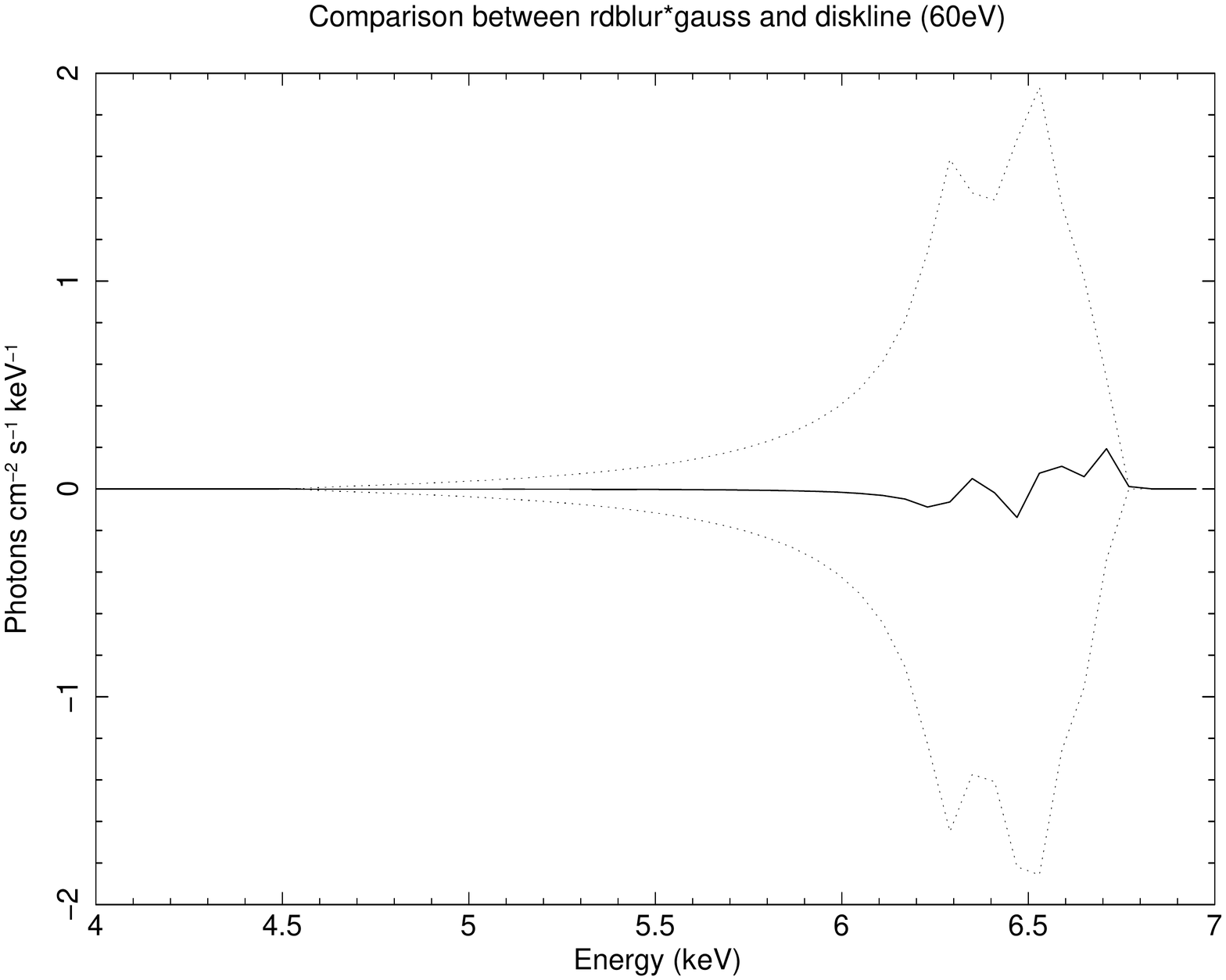}{0.6\textwidth}{(a)}
		\hspace{-1.5cm}
		\fig{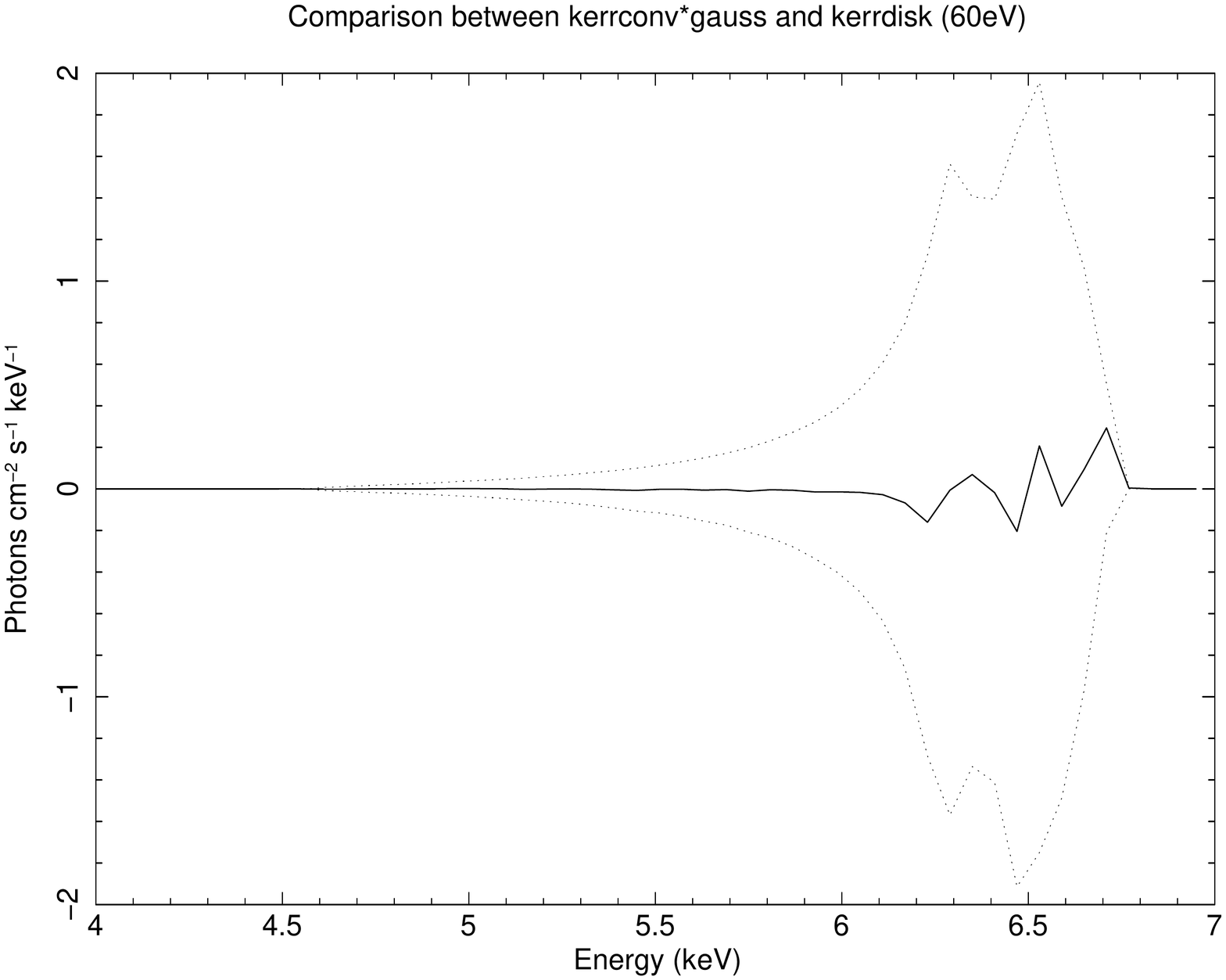}{0.6\textwidth}{(b)}
	}
	\caption{Comparisons, within \xspec, of relativistic single-line-emission models and their convolution version applied to a single-bin Gaussian, respectively with negative and positive normalizations. The difference between the profiles in each couple is depicted by the solid black line, while the profiles themselves are represented by dotted lines. Panel \textit{a} shows the comparison between \rb \gauss\ and \dl\ while panel \textit{b} shows \kec \gauss\ and \ked. In both panels, the energy grid over which the models are calculated is composed of 60-eV-wide bins.}
	\label{fig:comp1_60}
	
\end{figure}

\texttt{shadconv} calls its single-line counterpart and convolves it over the Gaussian profile through \xspec's own convolution routine, \conln, which uses fast Fourier transforms (FFTs) and can be found among the numerical utilities in standard \xspec\ builds.\footnote{As from at least \xspec\ 12.10.1m, it is defined in Xspec/src/XSUtil/Numerics/Convolution.h} \conln\ is also used (that we know of) by four models within \xspec, namely \texttt{kdblur}, \texttt{kdblur2}, \texttt{rdblur} and \texttt{kerrconv}, which constitute the convolution counterparts to the \texttt{laor}, \texttt{laor2}, \dl, and \ked\ models respectively. We therefore ran comparisons between these couples of models within \xspec, finding that they show the same behaviour we had encountered with our local model: all the above-mentioned comparisons were carried out by commands such as, e.g., \texttt{model rdblur*gaussian + diskline} and setting the normalizations to opposite values (1 for the \texttt{gaussian} component and -1 for the single-line profile, \dl\ in this example). Panels \textit{a} and \textit{b} of Fig. \ref{fig:comp1_60} show the results of comparing \dl\ to \rb\ and \ked\ to \kec\ as examples: the energy binning is set to 60 eV (obtained by using the command \texttt{energies 1.4 50.0 810}) and the parameters are set to the values in Table \ref{tab:pars} in all cases.  

\begin{figure}[h]
	\centering
	\gridline{
		\hspace{-1.cm}
		\fig{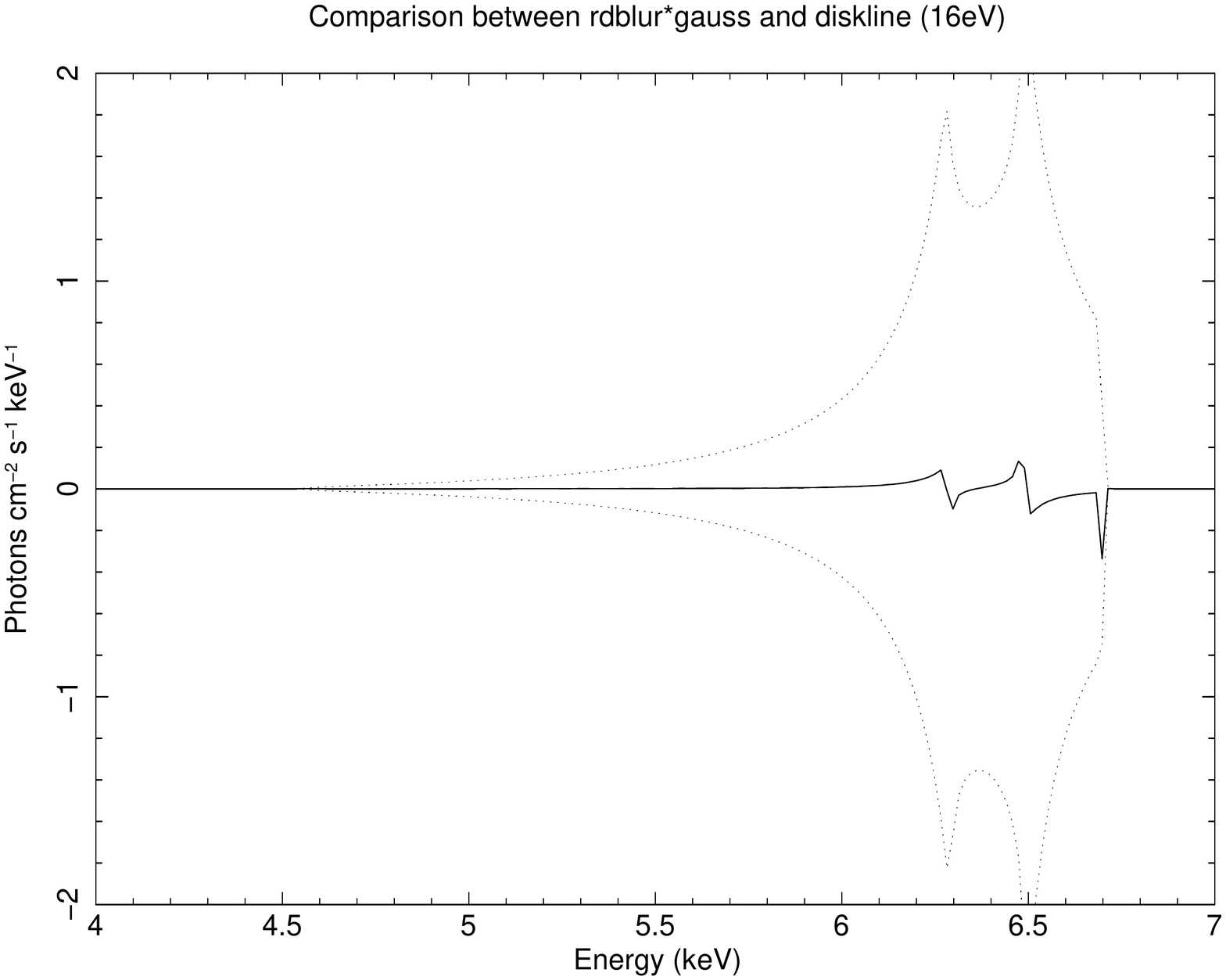}{0.6\textwidth}{(a)}
		\hspace{-1.5cm}
		\fig{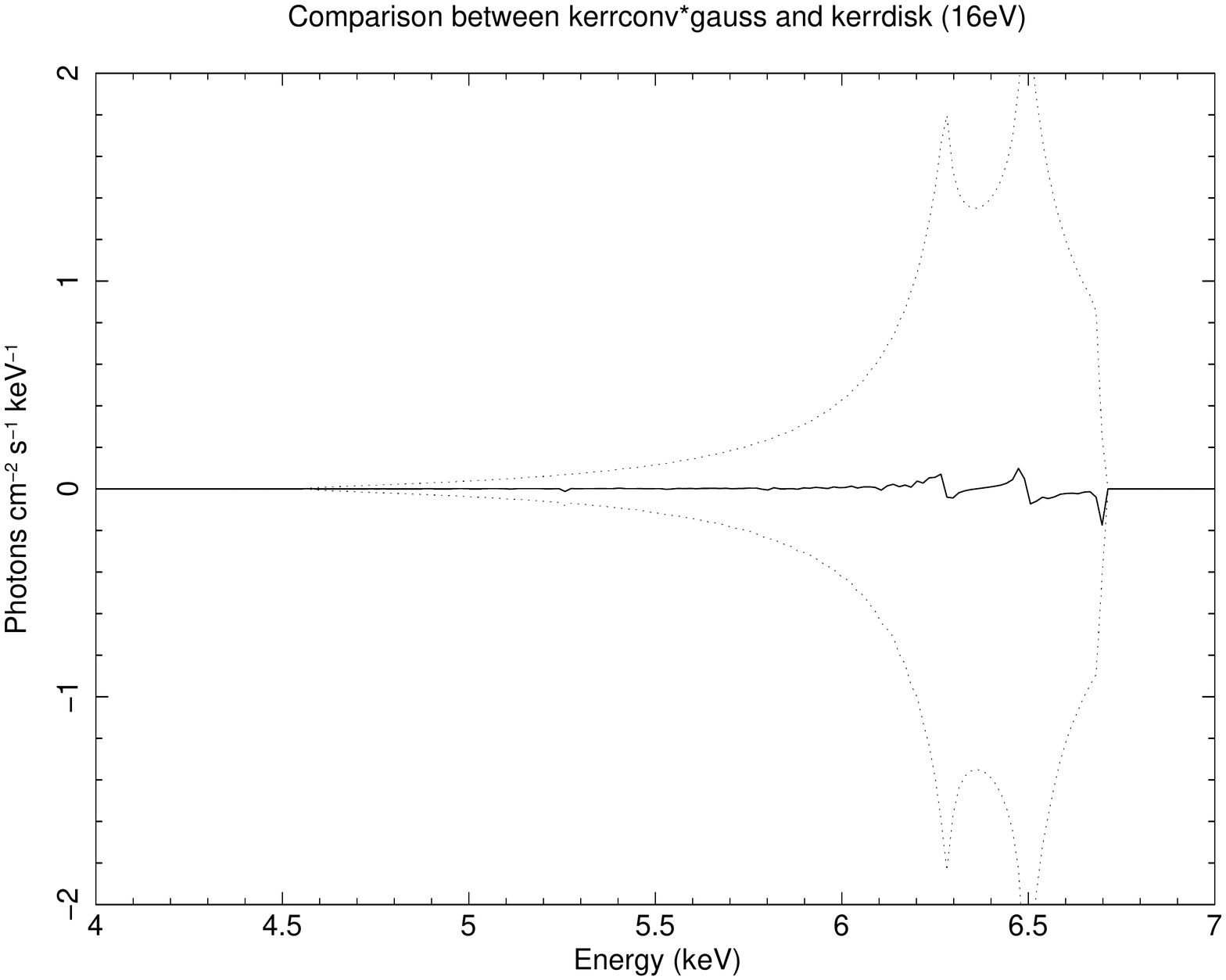}{0.6\textwidth}{(b)}
	}
	\caption{Comparisons, within \xspec, of relativistic single-line-emission models and their convolution version applied to a single-bin Gaussian, respectively with negative and positive normalizations. The difference between the profiles in each couple is depicted by the solid black line, while the profiles themselves are represented by dotted lines. Panel \textit{a} shows the comparison between \rb \gauss\ and \dl\ while panel \textit{b} shows \kec \gauss\ and \ked. In both panels, the energy grid over which the models are calculated  is composed of 16-eV-wide bins.}
	\label{fig:comp1_16}
	
\end{figure}

\begin{figure}[h]
	\centering
	\gridline{
		\hspace{-1.cm}
		\fig{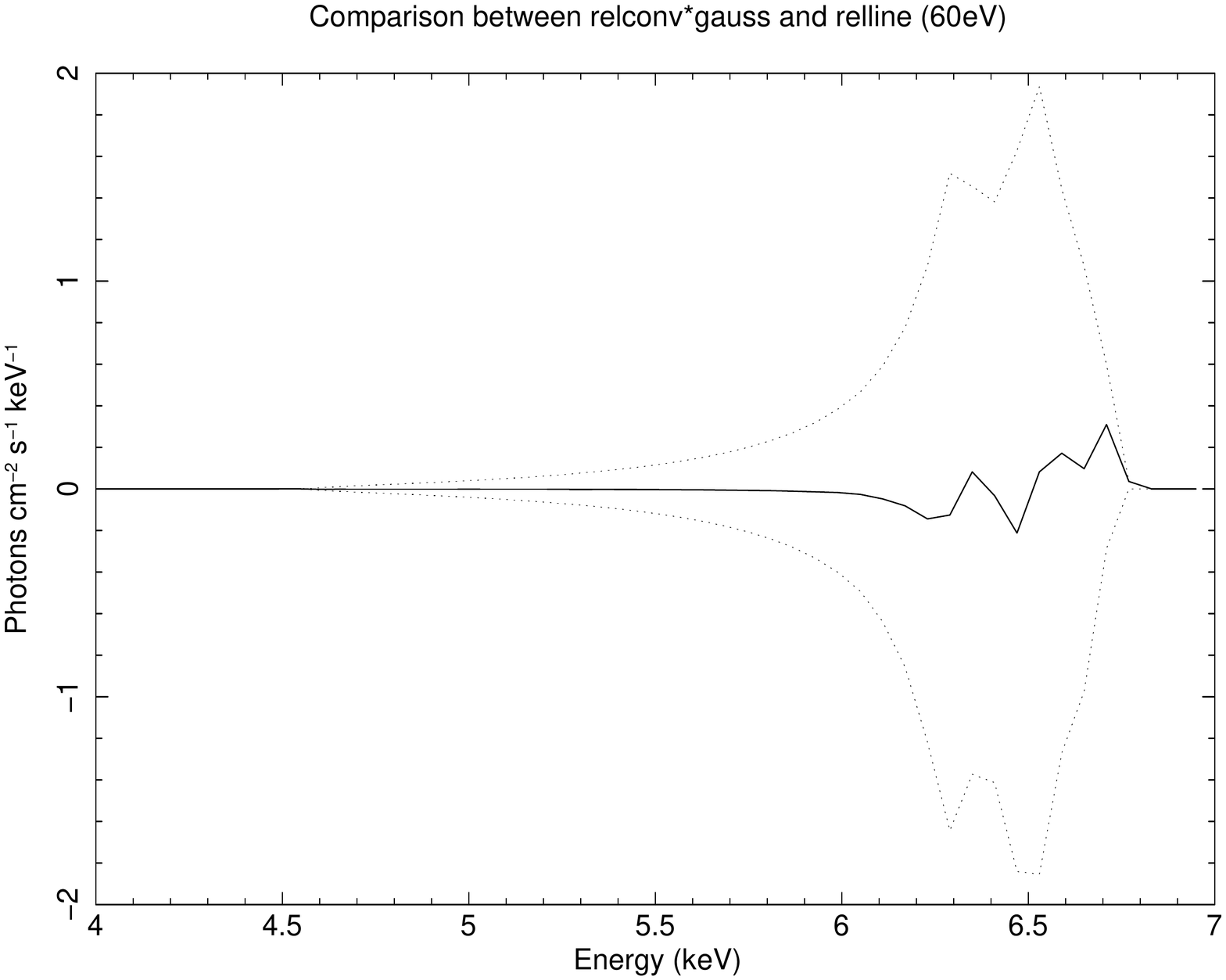}{0.6\textwidth}{(a)}
		\hspace{-1.5cm}
		\fig{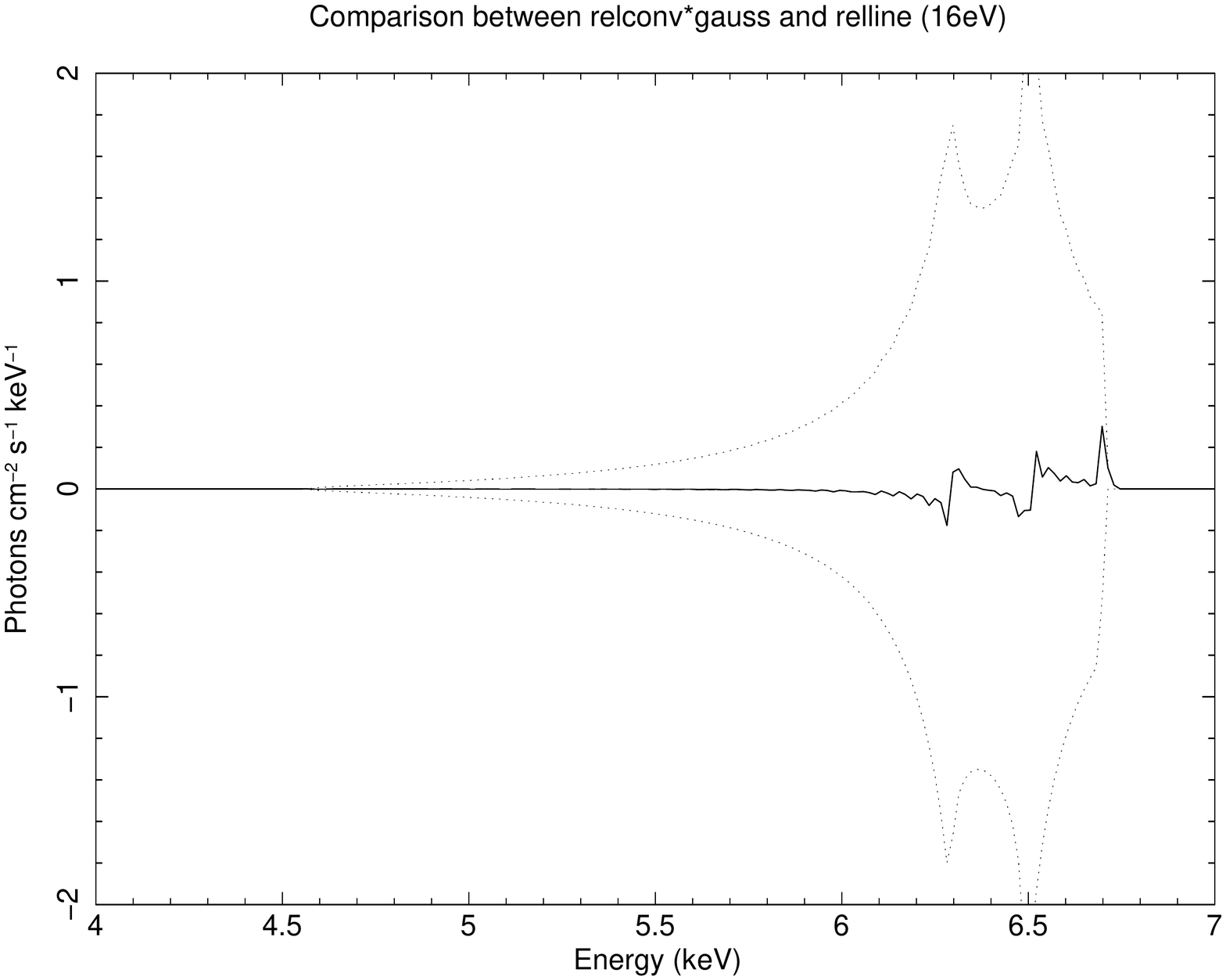}{0.6\textwidth}{(b)}
	}
	\caption{Comparisons, within \xspec, of line profiles generated with \rl\ and its convolution version, \rc, applied to a single-bin Gaussian, respectively with negative and positive normalizations. The difference between the profiles in each couple is depicted by the solid black line, while the profiles themselves are represented by dotted lines. The energy bins over which the models are computed are 60-eV wide in panel \textit{a} and 16-eV wide in panel \textit{b}.}
	\label{fig:comprel}
	
\end{figure}

The shape and size of these residuals can vary greatly with the shape of the line profile and with the energy binning chosen, but their local (i.e., in a given energy bin) value relative to the local flux can even exceed $ 10\% $, which is of the same order of or even larger than some minor effects currently studied in X-ray-emitting systems. Slow-changing profiles and denser energy grids show smaller residuals overall: see e.g. Fig. \ref{fig:comp1_16}, which uses 16-eV energy bins obtained through the command \texttt{energies 0.002 65.538 4096}. Interestingly enough, even other convolution models which do not rely on \conln\ show the same behaviour for the same parameters and energy grid (see e.g. the \rl\ case in Fig. \ref{fig:comprel}). As can be seen from the 16-eV binning cases, using a number of bins which is a power of two does not remove this effect, suggesting it is not due to problems in assigning the proper length to flux arrays for them to be used in FFTs.

\begin{figure}[h]
	\centering
	\gridline{
		\hspace{-0.5cm}
		\fig{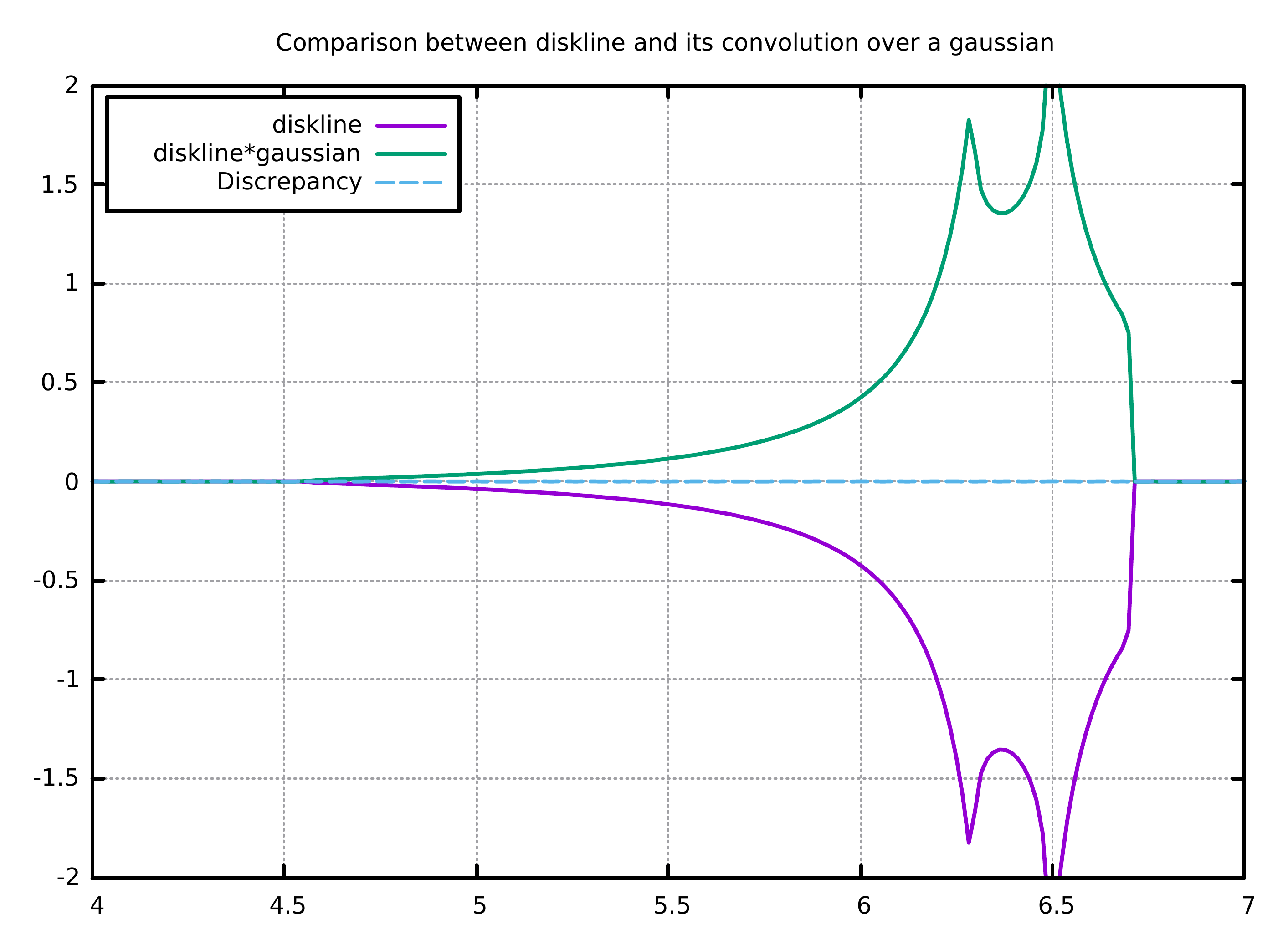}{0.5\textwidth}{(a)}
		\hspace{-0.0cm}
		\fig{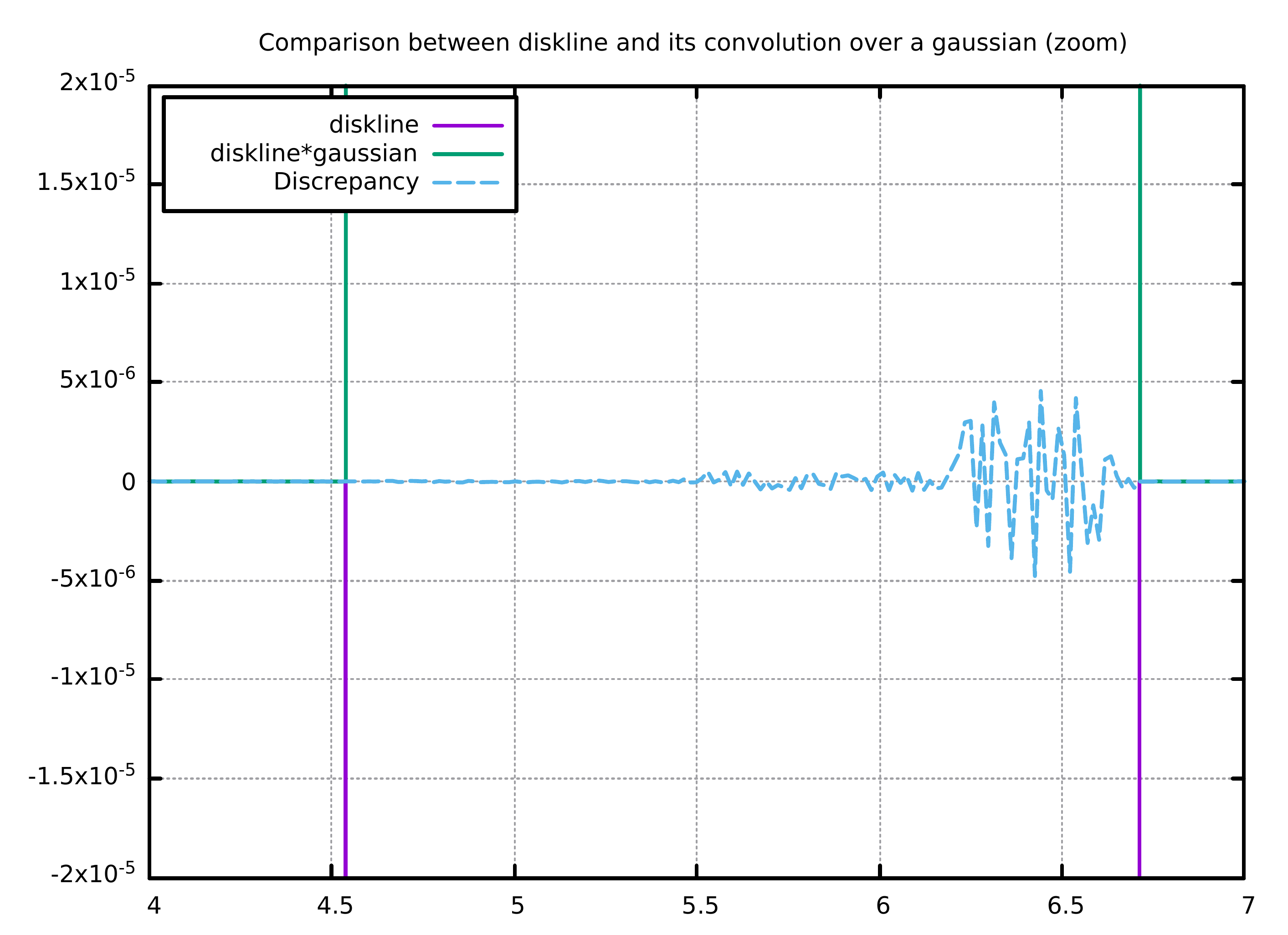}{0.5\textwidth}{(b)}
	}
	\caption{Panel \textit{a} shows the comparison of the line profile generated with \dl\ and the one obtained convolving it externally with a single-bin Gaussian with \texttt{NR}, respectively with negative and positive normalizations. The difference between the profiles is depicted by the dashed cyan line, while the profiles themselves are represented by solid lines. Panel \textit{b} shows a close-up in flux of the previous graph, on a scale five order of magnitudes lower than the one in panel \textit{a}, to make the small amount of residual noise visible. The energy grid in both panels is composed of 16-eV-wide bins.}
	\label{fig:comp_NR_proper}
	
\end{figure}

Running the convolution between any line profile and a single-bin Gaussian profile with external routines seems to show no such effects: we wrote a simple program by drawing on the convolution routines in the widely known Numerical Recipes in Fortran programming book and ran some tests convolving the same \xspec-generated profiles and single-bin Gaussians externally with it. Fig. \ref{fig:comp_NR_proper} shows an example of the virtually non-existent residuals in these comparisons. This external program, which we will call \texttt{NR} for the sake of brevity, also employs FFTs to carry out the convolution between the two profiles swiftly.


We strongly urge anyone working on convolution routines to be highly wary of these problems, and hope that the information here presented may at least point us all in the right direction towards solving them.  

\software{\dl\ \citep{diskline},
	\ked\ \citep{kerrdisk},	
	\texttt{laor} \citep{laor},
	Numerical Recipes in FORTRAN \citep{numerecipes},
	\rl\ \citep{relline},
	\sd\ \citep{shaddisk},
    \xspec\ \citep{xspec}}





\begin{table}
	\centering
	\caption{Model parameters used when generating all the line profiles shown in this note. Since \ked\ and \rl, and by extension \kec\ and \rc, allow more parameters than \dl, the spin and redshift in those models were set to zero. \ked\ and \kec\ accept radii only in units of the radius of the innermost stable circular orbit, which lies at $6~r_{g}$ in the Schwarzschild metric, therefore the values of $Rin_{ms}$ and $Rout_{ms}$ were set to 1.66667 and 166.667 respectively. The angular emissivity is considered isotropic in the case of \rl\ and \rc.}
	\begin{tabular}[c]{lLC}
		\hline
		Models   & $ Parameters $ & $Values $ \\
		\hline 
		\rb      & Betor10    &  -2     \\
		& Rin_M     &  10               \\          
		& Rout_M    &  1000          \\    
		& Incl &  30           \\   
		\texttt{gaussian}     & LineE & 6.4      \\
		& Sigma    &  0.0          \\ 
		& norm  &  1.0           \\  
		\dl    & LineE\ (\mathrm{keV})     &  6.4  \\ 
		& Betor10    &  -2     \\
		& Rin_M     &  10               \\          
		& Rout_M    &  1000          \\    
		& Incl &  30           \\    
		& norm  &  -1.0   \\          
		\hline 
	\end{tabular}
	\label{tab:pars}
\end{table}





\end{document}